\documentclass{osa-article}
\usepackage{graphicx}                 % inkludering av grafikk
\usepackage{xcolor}
\usepackage{subcaption}
\usepackage{amsmath,amsfonts,amssymb} % matematikksymboler
\journal{osac}
\articletype{Research Article}
\begin{document}

\title{Towards High Throughput Large Area Metalens Fabrication using UV-Nanoimprint lithography and Bosch Deep Reactive Ion Etching}

\author{Christopher A. Dirdal\authormark{*}, Geir Uri Jensen, Hallvard Angelskår, Paul Conrad Vaagen Thrane, Jo Gjessing,  Daniel Alfred Ordnung}

\address{\authormark{1}SINTEF Microsystems and Nanotechnology, Gaustadalleen 23C, 0737 Oslo, Norway}

% \author{Author One,\authormark{1} Author Two,\authormark{2,*} and Author Three\authormark{2,3}}

% \address{\authormark{1}Peer Review, Publications Department, The Optical Society (OSA), 2010 Massachusetts Avenue NW, Washington, DC 20036, USA\\
% \authormark{2}Publications Department, The Optical Society (OSA), 2010 Massachusetts Avenue NW, Washington, DC 20036, USA\\
% \authormark{3}Currently with the Department of Electronic Journals, The Optical Society (OSA), 2010 Massachusetts Avenue NW, Washington, DC 20036, USA}

\email{\authormark{*}christopher.dirdal@sintef.no} %% email address is required

% \homepage{http:...} %% author's URL, if desired

%%%%%%%%%%%%%%%%%%% abstract %%%%%%%%%%%%%%%%
%% [use \begin{abstract*}...\end{abstract*} if exempt from copyright]

\begin{abstract}
We demonstrate the fabrication of diffraction-limited dielectric metasurface lenses for NIR by use of standard industrial high throughput silicon processing techniques: UV Nano Imprint Lithography (UV-NIL) combined with continuous Reactive Ion Etching (RIE) and pulsed Bosch Deep Reactive Ion Etching (DRIE). As the research field of metasurfaces moves towards applications these techniques are relevant as potential replacements of commonly used cost-intensive fabrication methods utilizing Electron Beam Lithography. We show that washboard-type sidewall surface roughness arising from the Bosch DRIE process can be compensated for in the design of the metasurface, without deteriorating lens quality. Particular attention is given to fabrication challenges that must be overcome towards high throughput production of relevance to commercial applications. Lens efficiencies are measured to be 30\% and 17\% at wavelengths $\lambda=1.55\mu$m and $\lambda=1.31\mu$m, respectively. A number of routes towards process optimization are proposed in relation to encountered challenges. 
\end{abstract}

%%%%%%%%%%%%%%%%%%%%%%%%%%  body  %%%%%%%%%%%%%%%%%%%%%%%%%%

%\input{Introduction.tex}
%\input{Theory.tex}
%\input{Results.tex}
%\input{Discussion.tex}
%\input{Conclusion.tex}

\section{Introduction}
\subsection{Challenging the paradigm for optical sensor design}
The current revolution in sensor technologies is opening up for a wide number of new applications where optical components are required to be \textit{small, lightweight and cheap}, without compromise on optical quality. Relevant application areas include in-vivo medical imaging, drone-based imaging systems, mobile phones and wearables. A significant drawback for optical sensor technology in this context, however, is the fact that optical systems are generally \textit{big, heavy, and expensive}.% in contrast to electronic sensor systems which are generally small, lightweight and cheap. 
The recent developments within nanopatterning techniques and simulation tools have lead to the development of the research field known as \emph{metasurfaces} which may challenge this paradigm. For instance, the first proof-of-concepts have been published which show that metasurfaces can be used to move powerful microscopy techniques (which often require large table-mounted equipment) into the body. The authors of \cite{arbabi2018two}, demonstrate how dual-wavelength metasurface lenses can help to miniaturize two-photon imaging for e.g. in-vivo brain imaging, achieving comparable resolution to that of a conventional objective for a table top microscope. The authors of \cite{pahlevaninezhad2018nano} demonstrate superior resolution for their in-vivo optical coherence tomography relying on a metasurface lens. It is easy to imagine several other application areas where metasurfaces can make a significant change, e.g. miniaturizing hyperspectral or 3D imaging systems (which too can be large) so that they may be placed onto drones. Or how about spectrometers in cellular phones, or a holographic display on your clock?  

Metasurfaces are able to overcome the size, weight and cost constraints facing current optical sensor systems by allowing to fabricate optics using the same standard silicon (Si) processing technology used to fabricate electronics. In contrast to optical sensor systems, electronic sensor systems are generally small, lightweight and cheap. Although there currently do exist lithographical methods to make e.g. high quality curved microlenses, metasurfaces offer the advantage of being able to integrate a multitude of optical functions (e.g. lens, filter, polarizer) into a single surface. In this respect, metasurfaces have many similarities with diffractive optical elements. However, by utilizing optical resonances in nanostructures such as pillars, bricks or discs (rather than e.g. stepped gratings) metasurfaces offer unprecedented control over all degrees of freedom of the propagating field: The phase, intensity, polarization and dispersion. Furthermore metasurfaces can potentially be integrated into the same Si process lines which already are used for making e.g. detectors. This is a development with significant potential to save costs and reduce sizes, as microlenses and detectors currently rely on separate manufacturing lines in general.  

\subsection{Towards high throughput, large area patterning} \label{sec:TowardsHighThroughput}
As the research field has until now been primarily interested in demonstrating the potential of metasurfaces, most dielectric metasurface lenses (or \emph{meta}lenses) are fabricated by using the best resolution nanopatterning methodologies, despite tending to be slow and costly. To be more specific, virtually every paper on state-of-the art dielectric metalenses to date has relied on Electron Beam Lithography (EBL) \cite{kamali2016highly,arbabi2016miniature,arbabi2018two,arbabi2018mems, pahlevaninezhad2018nano,devlin2016high,khorasaninejad2016metalenses, chen2018broadband, chen2019broadband}. Here EBL is typically used in one of two ways: (i) EBL is used to pattern resist for a metal lift-off, thereby attaining a hard mask for subsequent etching of the (typically silicon) metasurface structures (typically for operation in NIR, but also VIS) \cite{kamali2016highly,arbabi2016miniature,arbabi2018two,arbabi2018mems,pahlevaninezhad2018nano}, or (ii) EBL is used to pattern high aspect ratio resist (as much as 15:1) holes for subsequent ALD deposition of TiO$_2$ which, after lift-off, yield the metasurface structures (typically for operation in VIS)  \cite{devlin2016high,khorasaninejad2016metalenses, chen2018broadband, chen2019broadband}). The latter technique is typically used when extreme structural requirements apply, such as for minimum gaps between metasurface pillars being less than 20nm. %either in order to make a hard-mask for a top-down approach (e.g. \cite{khorasaninejad2014silicon,arbabi2016miniature}) or for ALD deposition (e.g.  \cite{devlin2016high,khorasaninejad2016metalenses}). 
Moving on towards applications, it is therefore necessary to develop low cost, high throughput, large area patterning methods (as agreed upon in \cite{su2018advances, urbas2016roadmap, zhang2016printed, hsiao2017fundamentals, bishop2018metamaterials, barcelo2016nanoimprint, park2019all}) which at the same time offer comparable reproducibility and resolution as to that of EBL.

Several fabrication methods relevant to large area patterning have been proposed and partially applied to metasurfaces\cite{bishop2018metamaterials, hsiao2017fundamentals}, including nanoimprint lithography \cite{makarov2017multifold, wang2017nanoimprinted,yao2016nanoimprint, chen2015large,fafarman2012chemically,ibbotson2015optical, wu2018moire, gao2014nanoimprinting, sharp2014negative, bergmair2011single,franklin2015polarization,lucas2008nanoimprint,rinnerbauer2015nanoimprinted,zhang2017printed,lee2018metasurface, briere2019semiconductors}, interference lithography \cite{zhang2015large}, plasmonic lithography \cite{luo2015fabrication,su2018advances}, immersion lithography \cite{hu2018demonstration}, deep-ultraviolet projection lithography \cite{park2019all}, pattern transfer \cite{kim2019facile, checcucci2019multifunctional}, additive manufacturing \cite{wu2019perspective}, self-assembly \cite{cai2019solution, fafarman2012chemically, kim2018chemically}, and associated high-throughput roll-to-roll and stepping processes\cite{she2018large}.  Examples of NIL applied to non-lensing metasurface applications include Mie-resonant holes and line-structures for photoluminescence enhancement control\cite{makarov2017multifold, wang2017nanoimprinted}, line structures for unidirectional transmission \cite{yao2016nanoimprint}, colloidal Au nanocrystals acting as quarter wave plates\cite{chen2015large} and chemically tailored dielectric-to-metal transition surfaces \cite{fafarman2012chemically}, metallic nano-woodpiles (Moir\'e patterns) for photonic crystal bandgaps \cite{ibbotson2015optical, wu2018moire}, metal-dielectric stacked fishnet structures for negative index metamaterials \cite{gao2014nanoimprinting, sharp2014negative, bergmair2011single}, plasmonic structures for active tuning colour \cite{franklin2015polarization}, localized surface plasmon resonance control\cite{lucas2008nanoimprint}, plasmonic photonic crystal lattice acting as a plasmonic absorber \cite{rinnerbauer2015nanoimprinted}, and line structures which act as cylindrical beam generators \cite{zhang2017printed}. 

Despite the wide variety of publications on nanoimprint lithography applied to metasurfaces just mentioned, there are to our knowledge only a few examples in which dielectric metalenses have been made using nanoimprint \cite{lee2018metasurface, briere2019semiconductors}. This is possibly explained by the challenges involved for etching quality structures with vertical sidewalls and aspect ratios ranging between 2:1 to 15:1. Also, as mentioned above, for demonstrations and "proof of principle" prototypes, the time required by direct writing methods such as Electron Beam Lithography (EBL) is not critical. Nevertheless, transitioning into technological applications, this challenge must be addressed. The authors of \cite{briere2019semiconductors} found that using a classical parallel-plate Reactive Ion Etch (RIE) with a metallic mask yielded slanted sidewalls in the metasurface structures, which in turn seem likely to have reduced the optical quality of their lens. An alternative approach based on selective area sublimation was used to overcome this issue (but which is only applicable to crystalline materials). In \cite{lee2018metasurface} metalenses of good optical quality are reported, fabricated by evaporating a stack of SiO$_2$, Cr and Au onto a polymer stamp, after which the stack is transferred to a Si film on quartz substrate by imprinting. The de-attached SiO$_2$-Cr-Au stack is then used as an etch mask for the Si film. This method has the advantage of avoiding the need to pattern the hard mask through etching, but it seems likely that the polymer stamp must be cleaned or re-created for every imprint.

\subsection{The Bosch process in comparison to competing etching techniques}
The selection of the most appropriate plasma etch type for industrial metalens fabrication is not
clear-cut. One group of process alternatives is a continuous reactive ion etch (RIE) - be it a
classical parallel-plate RIE, or more advanced and better controlled inductively coupled plasma
(ICP) based RIE, or a capacitively coupled plasma (CCP) RIE. The most advanced etchers
are ICP-based. Another dry etch type is so-called cryogenic etch or cryo etch, which runs at
temperatures lower than minus 100$^\text{o}$C, also in a continuous fashion. The pulsed Bosch-type
process (with two pulses, or the extended Bosch with three pulses, for each etch step) is the third
of the main categories/candidates. Bosch deep reactive ion etch (DRIE) produces sidewalls that
are not formally straight, but indented with "scallops", which is the main feature distinguishing
Bosch from the others in terms of wall appearance. The "envelope" wall can be made very
close to vertical, though, and the scallops could be made as small as 10 nm (depending on mask
thickness and selectivity).

Cryo etch has experienced a certain popularity in R\&D - in particular due to its smooth and
mirror-like sidewalls and capability of high aspect ratio (HAR) etching. Cryo etch has, however,
been little used by industry, owing to its rather serious drawbacks - all stemming from a very high
demand on accurate temperature control of the wafer and its etched structures. This translates
into a lack of process controllability, uniformity, and repeatability, as well as the need for a
continuously running line for substrate/wafer cooling by liquid nitrogen. For HAR, Bosch as
well as Cryo can go much further than non-cryo continuous RIE alternatives, and Bosch is the
HAR dry etch process of choice in industry. Some recent indications exist that Cryo is gaining
increased interest also in industry \cite{Cryo2020} due to its specific merits compared with Bosch. One
merit of interest for this paper is the entirely smooth walls, which are preferable in masters for
nanoimprint lithography. Even a wall angle slightly lower than 90 degrees is preferred, and easier
made by cryo than Bosch.

For dielectric metalens structures, published papers show requirements on etch aspect ratios (ARs) ranging all the way from 2:1 to above 30:1 (see e.g. \cite{arbabi2018two,khorasaninejad2014silicon,khorasaninejad2016metalenses,chen2018broadband,kamali2016highly}, although \cite{khorasaninejad2014silicon} uses the structures for beam-splitting rather than lensing). Furthermore, the minimum gaps between neighboring pillars range from less than 20 nm to several hundreds of nm. These widely differing ranges stem from a combination of the wavelength of the application, other parts of the specification, and the applied technology and materials. Roughly speaking, dry etched silicon metastructures operating in NIR tend towards lower AR ranges \cite{arbabi2018two,khorasaninejad2014silicon}, whereas ALD deposited TiO$_2$ metastructures operating in VIS tend towards high aspect ratios (HAR) \cite{khorasaninejad2016metalenses, chen2018broadband, chen2019broadband} (if they instead were to be made by etching). In terms of selecting the most appropriate etch type, one should – perhaps a bit simplisticly - distinguish between low-to-medium range ARs, and a high aspect ratio (HAR) range. No strictly defined border exists between the two, and indeed it depends on several parameters and on one's final target, but a border could arguably lie very roughly at 10:1, or in some cases quite a bit higher. For low-to-medium ARs it is not always evident that a classical RIE or ICP-RIE (or CCP-RIE) must yield to a Bosch or a cryo etch, despite the latter two being clearly better than the others at HARs. Indeed, ref \cite{kamali2016highly} achieves etching of ARs of 9:1 by ICP-RIE. 

Still, continuous RIE could be more challenging than a Bosch or cryo process in obtaining straight (vertical) sidewalls (see e.g. \cite{khorasaninejad2014silicon}). Furthermore, a Bosch process stops more abruptly on a buried oxide layer (BOX), as provided by Silicon On Insulator (SOI) wafers, a convenient feature for precise height control. However, with the extremely small sideways dimensions in such metalens pillars (e.g., a pillar width of 55 nm in \cite{khorasaninejad2014silicon} and even 40 nm in the TiO2 case of \cite{khorasaninejad2016metalenses}), a very strict control of scallop size as well as sideways
"notching" (a badly controlled sideways etch that can appear due to charging of the oxide) is
required. The undesired notching effect could be mitigated by time-based stop of the Bosch
DRIE just before the BOX is reached, followed by a well-tuned continuous RIE step. Another possible argument against the Bosch process is that it will always result in a pillar wall shape defined by scallops. However, this paper will show that this effect by itself does not seriously deteriorate metalens performance when it is corrected for in the NIL master  - a key finding of our paper. 

As noted above, it is possible that for some metalens designs the distance between neighboring pillars could become seriously small; thus, a limit exists for how long one can compensate for scallops by making the master pillars wider. However, as scallops at least under some circumstances could be made as small as 10 nm, very little master correction may often be required. A published example \cite{khorasaninejad2016metalenses}, though, shows gaps smaller than 20 nm. This not only gives an AR of over 30:1 in their design, but also strains the viability of scallop correction. In such an extreme situation, a cryo etch may be the best process option – if it is available.

In terms of access, Bosch process equipment is currently much more available than cryo, in R\&D as well as industrial facilities. Almost all labs that do any serious silicon etching have Bosch processes at hand. However, the same basic plasma tool can be used for cryo as well as Bosch, with relatively limited alterations to enable cryo. It is thus probably more likely that for a metalens development project a Bosch rather than a cryo process would be used in an R\&D lab, while for an industrial enterprise one would think that comparative performance is the decisive factor.

All in all, there is in our opinion no clear and obvious "winner" in the etch type competition for metalenses. However, as long as the scallops of the Bosch-etched walls are not a serious hindrance performance-wise, and the pillar gaps are not extremely small combined with very high ARs, Bosch DRIE at the very least seems like a strong contender.

\subsection{Our contribution}
In this paper we present the utilization of standard industrial high throughput silicon processing techniques for the fabrication of diffraction-limited dielectric metasurface lenses for NIR: We have used UV Nano Imprint Lithography (UV-NIL) patterning of a resist mask with subsequent Continuous and Bosch Deep Reactive Ion Etching (DRIE) for fabricating quality high aspect ratio metastructures with vertical sidewalls. To our knowledge this is the first such demonstration of the combination of these techniques, which are highly relevant to the growing demand for developing high throughput, large area patterning techniques for dielectric metasurfaces. Furthermore, we present a detailed account of the processing steps and the challenges involved, in order to hopefully contribute to the advancement of UV-NIL and DRIE as a route to achieve this. Employing UV-NIL still requires the fabrication of a master wafer, typically using EBL, but the cost of this can be reduced by fabricating masters with single (or several) dies, which are replicated to pattern a full master wafer using stepper nanoimprint lithography (stepper NIL) and reactive ion etching. However, full-wafer patterning by stepper NIL is not addressed in this paper.

\section{Design of metalens} \label{sec:Design}

\subsection{Physical principle} \label{sec:PhysPrinc}
The optical design of the metasurfaces relies on dielectric rectangular pillar arrays (Fig. \ref{fig:SimStructures}) and the widely used geometric phase principle \cite{kang2012wave, khorasaninejad2016metalenses,wang2017broadband, chen2019broadband}.

\begin{figure}
    \centering
     \begin{subfigure}[b]{0.40\textwidth}
         \centering
         \includegraphics[width=\textwidth]{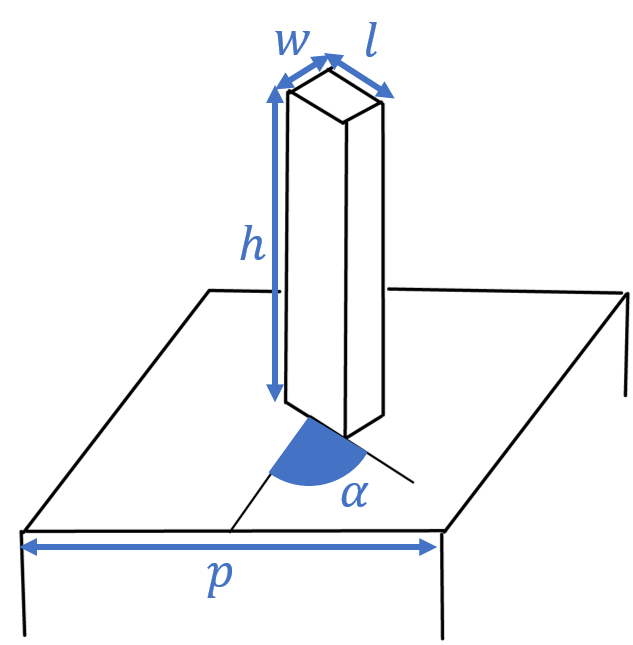}
         \caption{}
         \label{fig:UnitCellPic}
     \end{subfigure}
     \hfill
     \begin{subfigure}[b]{0.55\textwidth}
         \centering
         \includegraphics[width=\textwidth]{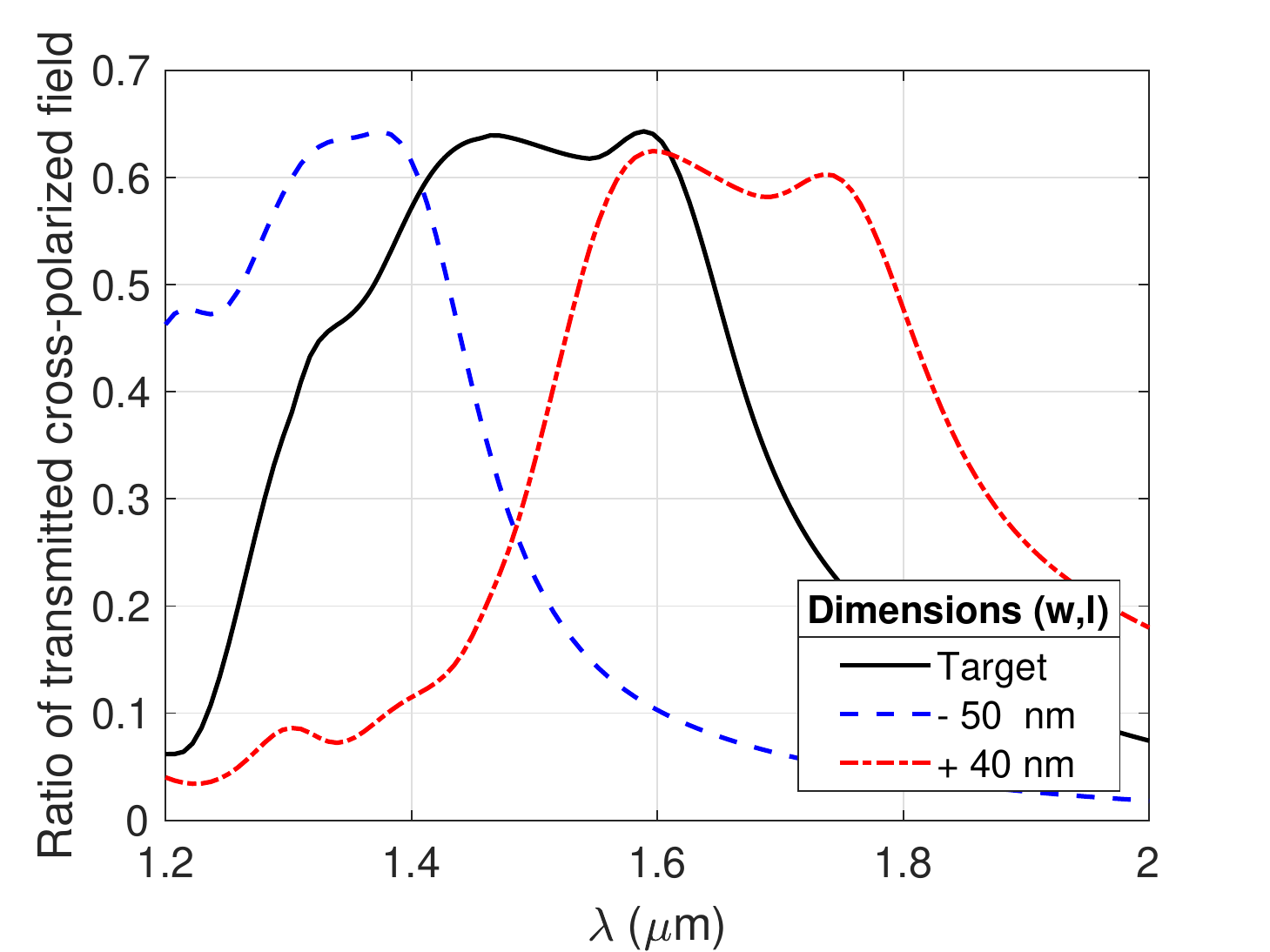}
         \caption{}
         \label{fig:StructuralTolerances}
     \end{subfigure}
    \caption{(a) Sketch of Si rectangular pillars rotated by an angle $\alpha$ relative to the unit cell axes. When the incident field is circularly polarized, the rotation angle imposes a phase shift of $2\alpha$ to the transmitted cross-polarized field. (b) Simulated cross-polarized intensity for left-circular field passing through an array of the sketched pillars using the Rigorously Coupled Wave Analysis (RCWA) method. The target dimensions of $h=1200$ nm, width $w=230$ nm, length $l=354$ nm and periodicity $p=835$ nm give the solid curve. Simulations for structures with reduced or increased lateral dimensions (by -50nm and +40nm) are displayed: These demonstrate that fabrication tolerances of at least $\pm 40$ nm in the lateral dimensions should give functioning metasurfaces at either of two common telecom wavelengths $\lambda = 1.55\mu$m or $\lambda = 1.31\mu$m.}
    \label{fig:my_label}
\end{figure}

The phase function $\phi(r)$ of a lens (which focuses normally incident light to a focal point a distance $f$ from the center of the lens) is given by
\begin{eqnarray} \label{eq:LensPhase}
\phi(r) = \frac{2\pi}{\lambda}\bigg (\sqrt{r^2 + f^2}-f \bigg ).
\end{eqnarray} where $\lambda$ is the wavelength of interest and $r$ is the radial distance from the center. The job of the metalens is to add the phase amount $\phi(r)$ to the incoming field at each point $r$ on the metasurface. If the incoming field is circularly polarized, phase can be added to the field by transmitting it through rotated rectangular pillars on the metasurface, rotated by an angle
\begin{eqnarray} \label{eq:Rotation}
\alpha(r) = \phi(r)/2,
\end{eqnarray}{} as sketched in Fig. \ref{fig:UnitCellPic}. This is known as the \emph{geometric phase principle}, in which the transmitted field $|E_\text{out}\rangle$ may be expressed as

{}

\begin{eqnarray} \label{eq:crossPol}
|E_\text{out}\rangle = \frac{t_x + t_y}{2}|L\rangle + \frac{t_x - t_y}{2} \exp(i\phi)|R\rangle, 
\end{eqnarray} where we have assumed that the incoming field is left handed circular polarized $|L\rangle$, and $|R\rangle$ is then the cross-polarized, right handed circular polarized field. $t_x$ and $t_y$ are the complex transmission coefficients for linear polarization directions orthogonal to the surface normal (along the coordinate $x$ and $y$ axes, respectively). Observing the transmitted field, it is clear that the values of the phase function \eqref{eq:LensPhase} are applied to the cross-polarized field $|R\rangle$ through the term $\exp(i\phi)$: I.e. the cross-polarized field will be focused to the focal point $f$, while the field remaining in the original polarization state is not. By appropriately designing the dielectric pillar periodicity $p$, height $h$, width $w$ and length $l$ one can tune $t_x$ and $t_y$ to increase the proportion of the transmitted field which is focused: By tuning the parameters such that $t_y=-t_x\equiv -t$ the metasurface also acts as a quarter-wave plate in which all the field is cross polarized, giving

\begin{eqnarray} \label{eq:crossPol-QWP}
|E_\text{out}\rangle = t\exp(i\phi)|R\rangle,
\end{eqnarray} where now all the transmitted field is focused at the focal point. Since the phase $\phi(r)$ is imposed through the rotation \eqref{eq:Rotation} alone, the simulation task is limited to finding the dimensions $p,h,w,l$ of the rectangular pillar array which optimize the degree of cross-polarization of the transmitted field. By fixing a common height $h$ for all of the pillars, the metasurface can be flat and well suited for fabrication using lateral patterning techniques. Furthermore, as is common in literature (e.g. \cite{kang2012wave,khorasaninejad2016metalenses}), we also apply the same values $w$ and $l$ to all rotated pillars and thereby disregard changes incurred upon $t_x$ and $t_y$ when rotating the rectangular pillars by the angle $\alpha$. This simplification allows us to use a continuous range of angles $\alpha \in [0,\pi)$ and  using identical (although rotated) pillars yields a constant filling factor over the UV-Nanoimprint Lithography stamp, which is an advantage towards process optimization (Sec. \ref{sec:NIL}). Preliminary simulations for the Si rectangular pillars on a quartz substrate indicate that the phase discrepancies incurred by this simplification are at most (varying by angle $\alpha$) on the order of around 0.03 rad. The transmittance discrepancies due to rotation seem to be negligible, however.

\subsection{Sweep simulations to find array dimensions} \label{sec:Simulations}
We performed sweep simulations to find array dimensions that maximize transmission of the cross polarized field using Rigorously Coupled Wave Analysis (RCWA) in the GD-Calc implementation and the Finite Difference Time Domain method (FDTD) in the OptiFDTD implementation. We find that dimensions of $h=1200$ nm, width $w=230$ nm, length $l=354$ nm and periodicity $p=835$ nm give full cross-polarization for the target wavelength of $\lambda = 1.55\mu$m. The simulations assume the source is placed within the silicon (Si) substrate: I.e. reflections at the wafer backside are neglected because they can be effectively eliminated by use of an anti-reflection coating, and are not intrinsic to the metasurface design. The ratio of transmitted cross-polarization intensity to the intensity of the light incident on the metasurface is shown in Fig. \ref{fig:StructuralTolerances}. The lower than unity ratio may be largely attributed to reflections at the boundary between the Si substrate ($n_\text{Si}=3.5$) and air ($n_\text{air}=1$): The Fresnel equations at normal incidence give roughly 31\% reflectance at a Si-air interface for the relevant wavelength bandwidth. The efficiency of the metalens can be increased by e.g. placing the Si metasurface pillars on a quartz substrate ($n_\text{SiO2}=1.5$) instead, which would reduce the corresponding reflectance to around 4\%. The structures on the interface may of course also contribute to reduce the expected efficiencies somewhat: Some scattering to diffraction orders within the Si substrate is expected since $\lambda/n_\text{Si}p=0.53$ for $\lambda=1.55\mu$m. %scattering to two higher orders $m=\pm 1$ within the Si substrate is permitted according to the grating equation.

Development of a UV-NIL and Bosch DRIE patterning process for metalens fabrication involves many parameters that must be taken into account when aiming to end up with the desired lateral dimensions found from simulations. As such it is useful to know what tolerances are permitted in the lateral dimensions of the structure, when planning for the fabrication. Figure \ref{fig:StructuralTolerances} shows two additional simulations where the lateral dimensions $w$ and $l$ of the pillars are varied to determine the permitted lateral fabrication tolerances. Increasing the lateral dimensions by 40 nm shows that the metasurface continues to have a high cross-polarization transmission at $\lambda=1.55\mu$m. While reducing the lateral dimensions by $-40$nm gives low cross-polarization transmission at $\lambda=1.55\mu$m, a high transmission is achieved at another common telecom wavelength of $\lambda=1.31\mu$m. Therefore, when given the freedom of using either $\lambda=1.55\mu$m or $\lambda=1.31\mu$m, the fabrication tolerance in the lateral dimensions is expected to be on the order of $\pm 40$nm. It is important to note that discrepancies in the lateral dimensions primarily affect the \emph{efficiency} of the lens and \emph{not} the focal spot size owing to the geometric phase effect (phase is imposed by rotation of the structure rather than its particular dimensions). This explains why our lenses fabricated in Sec. \ref{sec:results} attain diffracton-limited focusing despite slightly missing the target dimensions. Since high precision in reaching the target dimensions will be challenging under process development we have designed three designs to account for scenarios in which we might over- and under-estimate the end result dimensions. The variants of dimensions used for the fabrication of the NIL master are outlined in the table below.

\begin{table}[]
\centering
\caption{Lateral dimensions and filling factors of rectangular pillars for mask fabrication}
\label{tab:LateralDimensions}
\begin{tabular}{c|ccc}
Metasurface  & $w$ [nm]    & $l$ [nm] & Filling factor (F)        \\ \hline
A & 292 & 416 &  0.17\\
B & 237 & 361 &  0.12\\
C & 351 & 475 &  0.24
\end{tabular}
\end{table}It turned out that the smallest variant (i.e. variant B) of the table was the best suited due to broadening at the base of the resist pillars (as discussed in Sec. \ref{sec:NIL}). %In addition, adhesion issues seemed to be more of an issue for variants A and C.

\subsection{Compensation for Bosch sidewall surface roughness} \label{sec:ScallopCompensation}
\begin{figure}
     \centering
     \begin{subfigure}[b]{0.65\textwidth}
         \centering
         \includegraphics[width=\textwidth]{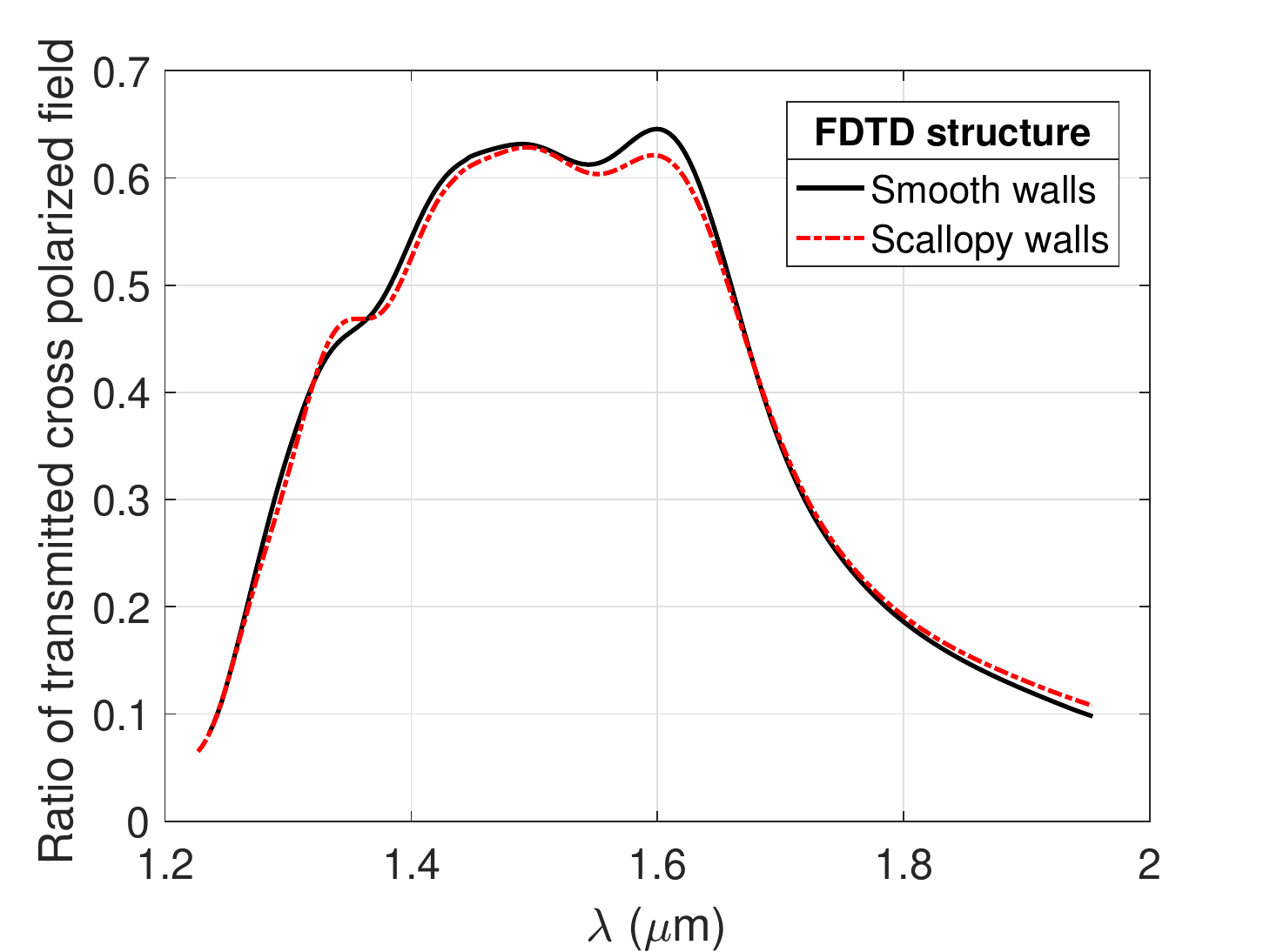}
         \caption{}
         \label{fig:SimPlot}
     \end{subfigure}
     \hfill
     \begin{subfigure}[b]{0.65\textwidth}
         \centering
         \includegraphics[width=\textwidth]{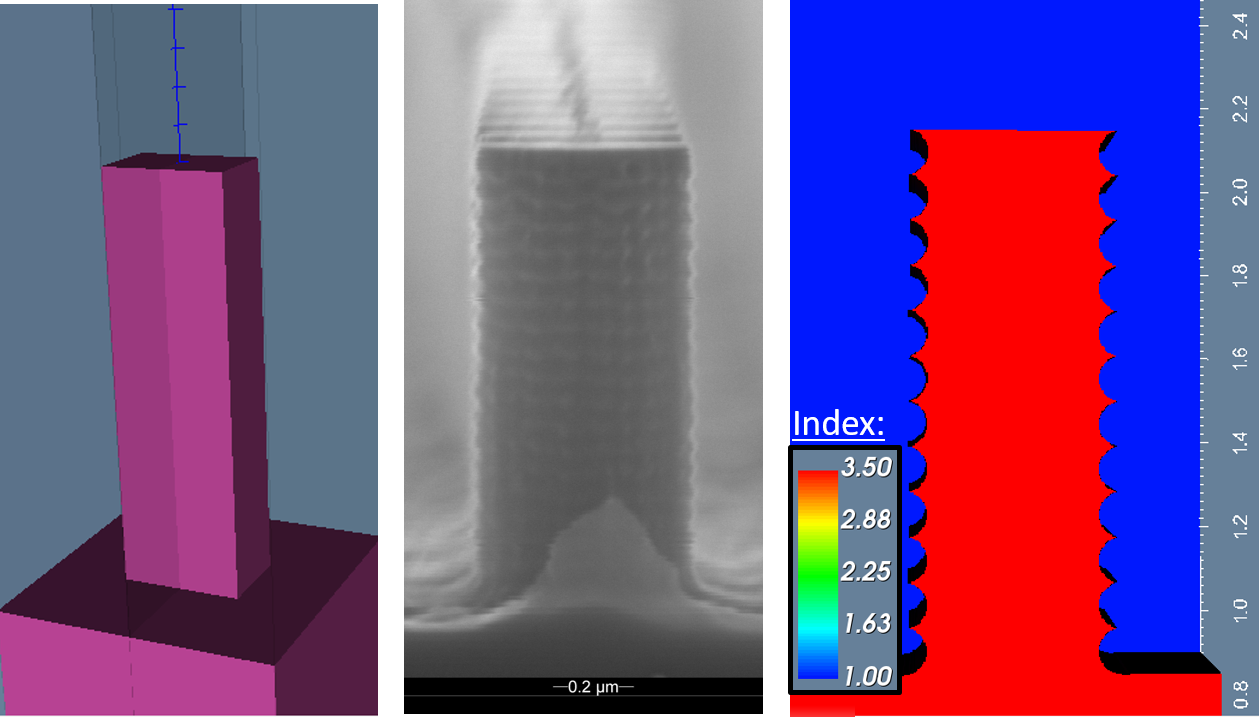}
         \caption{}
         \label{fig:SimStructures}
     \end{subfigure}
    %  \hfill
    %  \begin{subfigure}[b]{0.3\textwidth}
    %      \centering
    %      \includegraphics[width=\textwidth]{graph3}
    %      \caption{$y=5/x$}
    %      \label{fig:five over x}
    %  \end{subfigure}
\caption{(a) Simulated cross-polarized intensity for left-circular field passing through a metasurface consisting of rectangular silicon pillars ($n=3.5$) using the Finite Element Time Domain (FDTD) method. A FDTD simulation is shown also for a metasurface structure with washboard sidewall roughness ("scallopy") sidewalls, resembling the appearance after Bosch Deep Reactive Ion Etching (DRIE), in which the lateral dimensions of the rectangular pillar have been increased to compensate for the volume loss. The scallopy structure gives qualitatively equal results to the former FDTD simulation. (b) \emph{Left}: The structure with smooth sidewalls for the FDTD simulations, with height $h=1200$ nm, width $w=230$ nm, length $l=354$ nm and periodicity $p=835$ nm. \emph{Center}: A SEM image of a rectangular pillar etched into Si using Bosch DRIE. \emph{Right}: Showing the refractive index cross-sectional profile of the FDTD simulaton structure for an imitation of a rectangular Si pillar with "scallopy" sidewalls (scallop radii of $R=50$nm) where the lateral dimensions have been scaled up to $w'=320$ nm and $l'=449$nm in order to compensate for the volume loss by the scallops.}
\label{fig:Simulations}
\end{figure}

The center picture in Fig. \ref{fig:SimStructures} shows a SEM image of a Si rectangular pillar fabricated after Bosch-type 3-steps Deep Reactive Ion Etching (DRIE). As can be seen, the alternation of isotropic etching, passivation and de-passivation in the Bosch process leads to washboard surface patterns in the form of "scallops" which for simplicity have been characterized in terms of a scallop radius $R$. In the research field it is sometimes pointed out that surface roughness poses a problem towards achieving high optical quality\cite{khorasaninejad2016metalenses}, however since the roughness in this case is regular and occurs on length scales that are much smaller than the wavelength they can be treated as giving rise to effectively reduced dimensions which can be compensated for. The structure displayed on the right hand side of Fig. \ref{fig:SimStructures} shows a simulation model  mimicking a Bosch processed pillar using scallops with radius $R=50$ nm (as the one seen in the center of Fig. \ref{fig:SimStructures}). Fig. \ref{fig:SimPlot} presents the simulation labeled \textit{FDTD Scallopy} that shows that scallops can be compensated for by increasing the lateral dimensions corresponding to the volume loss represented by the scallops: Essentially the same transmitted cross-polarization is achieved by scaling the width and length according to

\begin{eqnarray}
w' &=& 1.0382 \bigg (w + \frac{\pi R}{2} \bigg ) = 320 \ \text{nm}, \\
l' &=& 1.0382 \bigg (l + \frac{\pi R}{2} \bigg ) = 449 \ \text{nm},
\end{eqnarray} which corresponds to a scaling that is $\sim 3.82\%$ larger than that required to compensate for the direct volume loss.

\section{Results} \label{sec:results}
This section describes the results of the UV-nanoimprint lithrography (UV-NIL) and etching steps, as well as some of the challenges encountered. Proposed strategies towards process optimization are discussed in Sec. \ref{sec:Discussion}.

\subsection{Imprint results} \label{sec:NIL}
\begin{figure}
     \centering
     \begin{subfigure}[b]{0.4\textwidth}
         \centering
         \includegraphics[width=\textwidth]{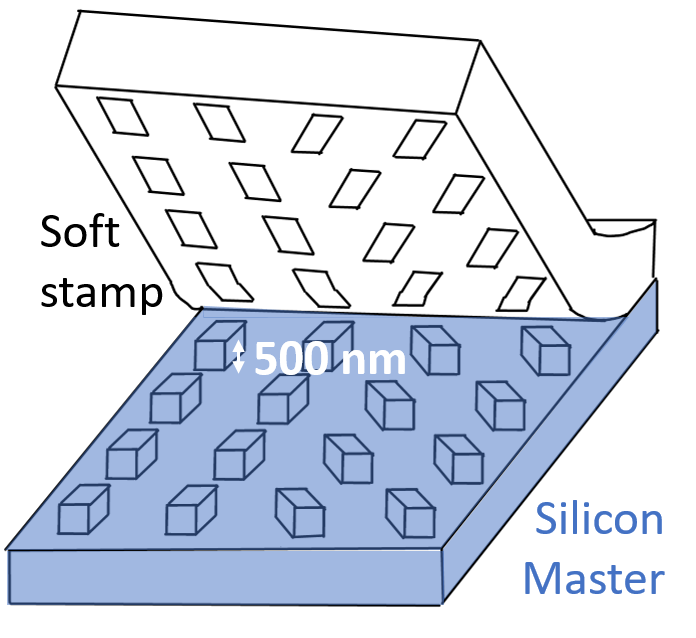}
         \caption{}
         \label{fig:Sketch_Softstamp}
     \end{subfigure}
    \ \
     \begin{subfigure}[b]{0.4\textwidth}
         \centering
         \includegraphics[width=\textwidth]{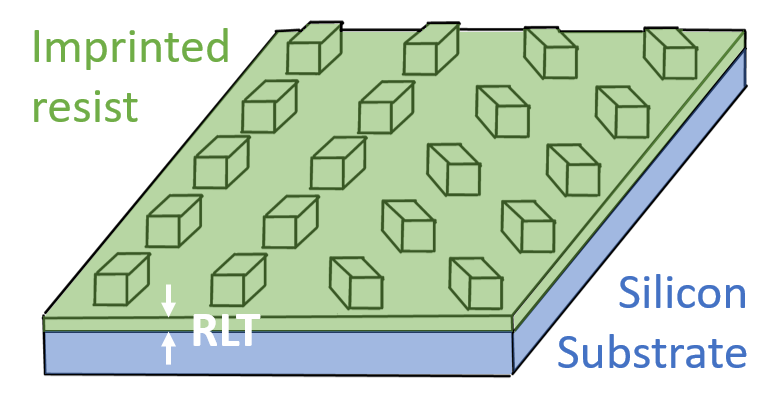}
         \caption{}
         \label{fig:Sketch_Imprint}
     \end{subfigure}
          \hfill
     \begin{subfigure}[b]{0.8\textwidth}
         \centering
         \includegraphics[width=\textwidth]{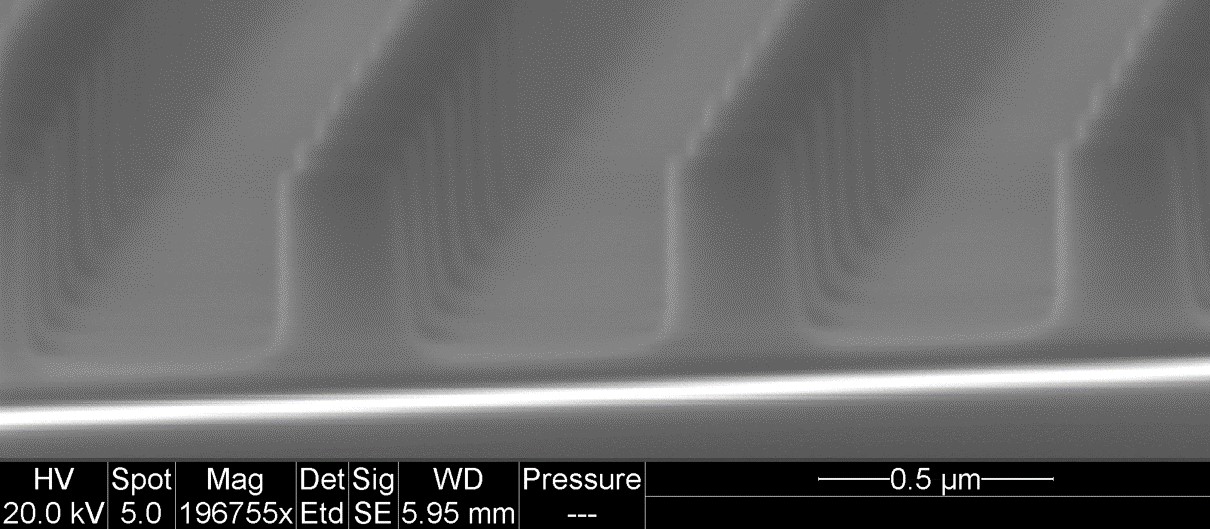}
         \caption{}
         \label{fig:ImprintmrNIL210}
     \end{subfigure}
        \caption{(a) \textit{Sketch of stamp fabrication}: A soft stamp is made by spinning Solvay Fomblin MD-40 onto a Si master wafer covered with nominally 500nm tall rectangular pillars, rolling on a carrier foil, exposure by UV illumination and finally detaching the stamp. The stamp is therefore an inverted copy of the silicon master, consisting of rectangular holes which are nominally 500nm deep. (b) \textit{Sketch of resist mask after imprint}: The soft stamp has been rolled onto a film of Micro Resist Technology mr-NIL210 series resist covering a bulk Si wafer substrate primed with mr-APS1. After exposure to UV illumination, the stamp is removed. What remains is patterned resist (a copy of the original master). Between the patterned resist and the silicon substrate there exists a film of residual resist characterized by a Residual Layer Thickness (RLT).  (c) \textit{Cross-sectional SEM image of imprinted and exposed resist on a silicon substrate}: The bright white line is probably caused by delaminated resist at the edge. Between the resist pillars and the silicon substrate one can observe the RLT. One also observes a broadening at the base of the resist pillar (in the shape of a "top-hat").}
        \label{fig:three graphs}
\end{figure}

UV-NIL was done using the Micro Resist Technology mrNIL-210 series resist and a soft stamp (Solvay Fomblin MD-40). The stamp is an inverted copy of a silicon master wafer with nominally 500 nm tall pillars forming the metasurface (see Fig. \ref{fig:Sketch_Softstamp}). In order to transfer this pattern to the silicon process wafers two conditions must be fulfilled: 1) the resist needs to be thick enough to function as an etch mask for more than 1.2 µm silicon DRIE, and 2) the residual layer thickness (RLT) of the remaining resist between the wafer surface and the imprinted pattern must be minimized (see Fig. \ref{fig:Sketch_Imprint}). In order to completely fill the inverted metasurface-structures in the stamp with resist, only a thin initial resist layer of less than 100 nm was needed for our design. However, in some cases it could be beneficial to have a thicker resist layer outside the patterned area, in order to prevent this area from being etched. Thus, resist films of different initial thicknesses from 500 nm and thinner were explored. The mrNIL210-200nm formulation, spun at 3000 rpm, gave low enough RLT values and acceptable variation over the metasurface. The resist thickness before imprint was measured by ellipsometry to be approximately 150 nm. The RLT obtained after imprint varied between $48$nm $\leq$ RLT $\leq$74 nm over the metasurface. While these RLT values were acceptable for further processing, the results obtained from the thicker mr-NIL210-500nm formulation were not viable for the metasurface patterning by DRIE: The RLT varied considerably over the lens, being at most always close to the pre-imprint resist film thickness. This finding turned out to be crucial for the ensuing fabrication.

\begin{figure}
    \centering
    \includegraphics[width=.65\linewidth]{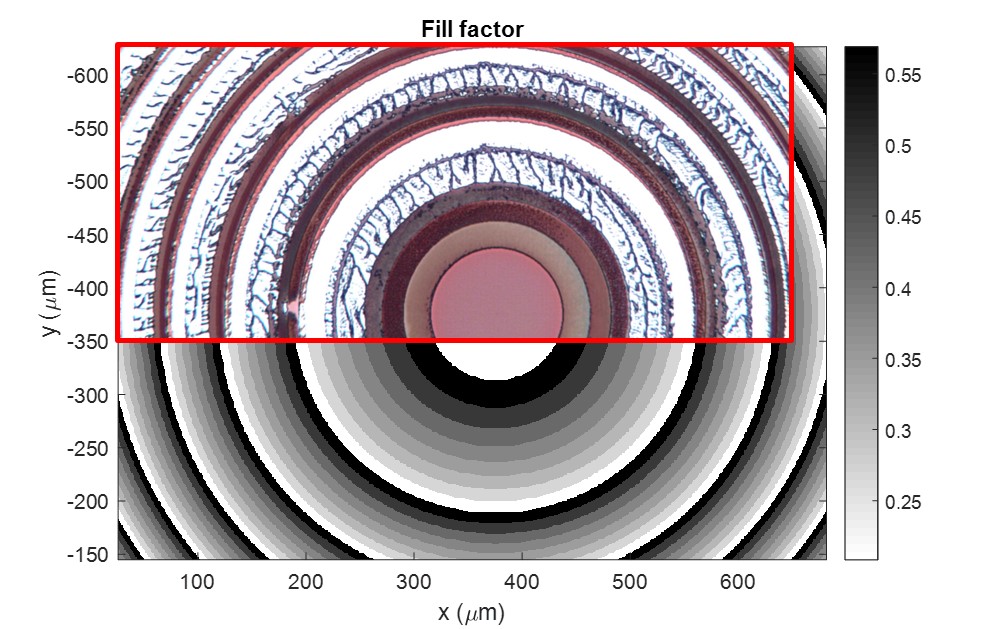}
    \caption{Imprint challenges when filling factor varies over metasurface. Here a cropped microscope image of imprinted resist (red border) is placed on top of the plotted fill factor (gray scale plot) of a metalens consisting of cylindrical pillars of varying radii. As can be seen, the structural fidelity of the imprint varies dramatically with filling factor (F): Areas with large F seem to turn out well, whereas areas with low F seem to detach with the stamp (apart from the center area).}
    \label{fig:RLTIssues}
\end{figure}

In general metasurfaces consist of structures of varying geometry, which means that the filling factor $F$ varies over the surface. Optimizing the RLT therefore becomes challenging, as the amount of resist used to fill the structures varies over metasurface. In this respect, our optical design based on the geometric phase method (Sec. \ref{sec:PhysPrinc}) has the advantage of providing identical structures (although rotated) with identical filling factors over the metasurface. This makes process optimization of the residual layer thickness easier. As a side remark: We also attempted fabrication of another optical design based on cylinders of varying radii in which issues with delamination of resist upon stamp removal seemed to depend on filling factor of the cylinders (see Fig. \ref{fig:RLTIssues}). 

An issue with the imprinted structures is broadening close to the base of the resist pillars (a resist "foot") as seen in Fig. \ref{fig:ImprintmrNIL210}. Such broadening is also frequently observed in SEM images from literature \cite{hamdana2018nanoindentation, si2017consecutive, plachetka2013tailored}. This resist foot leads to an added length in the lateral dimensions of the rectangular pillars, which is transferred to the final pillar dimensions in the patterned silicon (Sec. \ref{sec:BoschEtch}). We believe this broadening effect likely originates from the master wafer (from which the soft stamp is made) since the UV-cured resist generally follows the pattern of the master. Section \ref{sec:BoschEtch} outlines how we resolved the issue by means of various etch parameters for the resist pillars. 

Attaining sufficient adhesion between the resist and the substrate remains an important issue.
This is necessary to avoid delamination of the metastructure when the stamp is withdrawn from the surface after exposure. Such adhesion issues are unwanted if NIL is to become a high-throughput metasurface fabrication technique. To facilitate adhesion, RCA cleaned substrates were plasma activated (600W for 10mins) before spinning adhesion promoter (mr-APS1) immediately afterwards. Three different dimensions of resist pillars corresponding to filling factors $F_1=0.12$, $F_2 = 0.17$, $F_3 = 0.24$, respectively, were used for the metastructure (corresponding to the dimensions in Table \ref{tab:LateralDimensions}). However only the smallest gave reliable imprinting. The imprints with larger filling factors more or less consistently delaminated when withdrawing the stamp after exposure.

\subsection{Etch Methodology and Results} \label{sec:BoschEtch}
\begin{figure}
     \centering
     \begin{subfigure}[b]{0.44\textwidth}
         \centering
         \includegraphics[width=\textwidth]{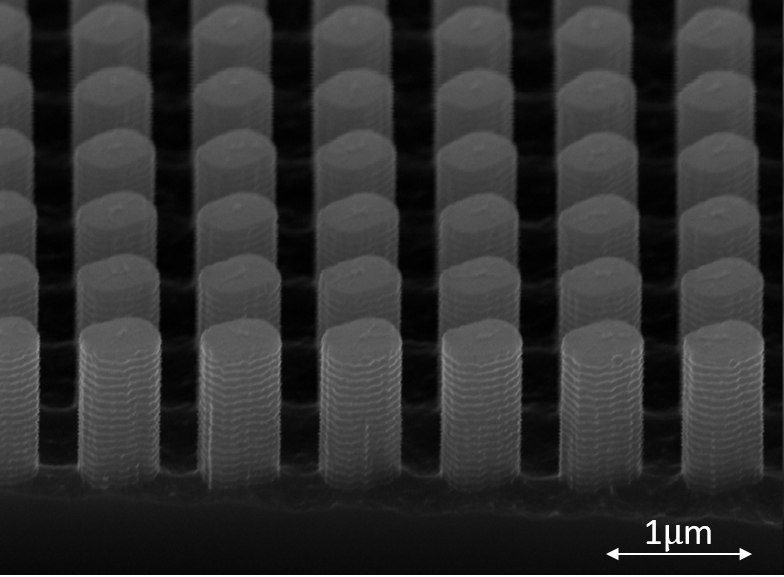}
         \caption{}
         \label{fig:RIEOverview}
     \end{subfigure}
     \hfill
     \begin{subfigure}[b]{0.52\textwidth}
         \centering
         \includegraphics[width=\textwidth]{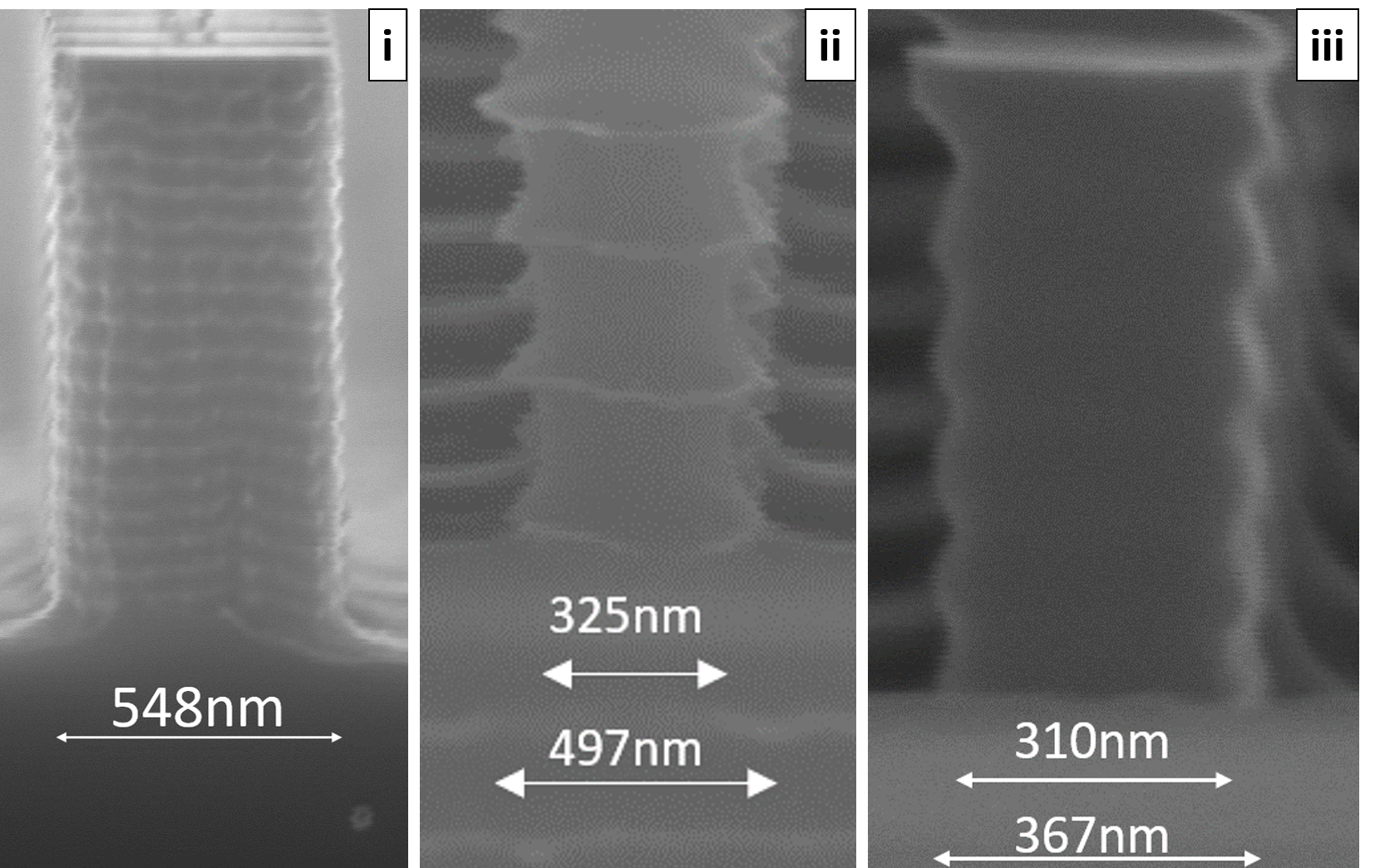}
         \caption{}
         \label{fig:ComparisonDRIE}
     \end{subfigure}
        \caption{(a) Patterned silicon after Bosch Deep Reactive Ion Etching (DRIE) of the silicon wafer with imprinted resist pictured in Fig. \ref{fig:ImprintmrNIL210}. The pulsed etching of the three-step Bosch process leads to washboard sidewall surface roughness. (b) Three  different scallop depths achieved using DRIE: (i) A close-up of the metasurface pictured in (a) with scallop depths of around $\sim 14$nm, (ii) In order to reduce the effective dimensions of the pillars, closer to the target dimensions, the scallop depths were increased to $\sim 86$ nm. Note that the resist has not been stripped in this image (although the resist is not clearly seen in the image). (iii) Dimensions close to the target were achieved by first performing a Bosch DRIE leading to scallops of depths $\sim 44$ nm, thereafter performing an oxidation step and oxide strip which in the end lead to scallop depths of $\sim 29$ nm.}
        \label{fig:RIEResults}
\end{figure}

In order to transfer the imprint patterns (as shown in Fig. \ref{fig:ImprintmrNIL210}) to the silicon wafer we utilized first a continuous (un-pulsed) RIE step to etch through the residual layer of resist before commencing with Bosch 3-step DRIE - i.e. pulsed etching consisting of the three steps passivation, de-passivation and isotropic SF$_6$-based silicon etch. Fig.~\ref{fig:RIEOverview} shows that high pattern fidelity is achieved in the silicon: We observe vertical sidewalls (indented with scallops, discussed below) for the pillars of around 1.2 $\mu$m height. In a separate run we observed the same pattern fidelity to at least $1.6 \ \mu$m etch depth. The cyclically-pulsed etching of the Bosch process leaves a washboard-like surface roughness characterized by a scallop depth which depends on the parameters of the Bosch process. For Fig. \ref{fig:RIEOverview} the scallop depths are  $\sim 14$ nm. In making these structures we used 6'' Si bulk wafers on which only a small area was patterned: 4 metalenses (rectangular pillars) of area 1.5mm $\times$ 1.5 mm, and one metalens (cylindrical pillars) of 0.75mm $\times$ 0.75mm. During the first RIE dry-etch step (for residue layer removal) the resist is completely removed from the surrounding wafer surface, resulting in an etch loading close to 100\% for the following Bosch DRIE step (to etch the Si pillars).  

As discussed in Sec. \ref{sec:NIL}, the broadening of the resist pillars seen in Fig. \ref{fig:ImprintmrNIL210} leads to added dimensions in the etched structures. The pillars shown in Fig. \ref{fig:RIEOverview} have lateral dimensions of around 420 nm $\times$ 530 nm, i.e. roughly 180nm too large in both directions in comparison to the simulation designs in Sec. \ref{sec:Simulations}. As a result the optical properties of this metasurface lens are poor. To solve this issue without redesigning the mask, three approaches were tested. First we attempted an increased length of the continuous dry-etch step to attempt to completely remove the resist "foot" at the base of the imprinted resist pillars. Although this somewhat deteriorated the quality of the imprinted pattern (turning the resist pillars into pyramids), this did not seem detrimental to the patterning of the Si pillars. We expect that further development of the process parameters of the continuous dry-etch step will likely remove the unwanted broadening  (as e.g. seen in \cite{hamdana2018nanoindentation} where the resist "feet" are removed completely while keeping vertical sidewalls in the resist), but the aforementioned run did not reduce it sufficiently. Our second approach was to dramatically increase the scallop depths to $\sim 86$ nm (see Fig. \ref{fig:ComparisonDRIE} ii) and the lateral dimensions were on the order of 307nm $\times $460nm (measured between the tops of the washboard pattern). 
A third approach consisted in realizing less extreme scallop depths of $\sim 44$ nm and thereafter oxidizing the structures so that around 100 nm oxide resulted. After stripping this oxide away (which on a planar silicon surface would have resulted in 44 nm reduced silicon thickness on each surface) the scallop depths were reduced to around $\sim 29$ nm (see Fig \ref{fig:ComparisonDRIE} iii) and the lateral dimensions were on the order of 210nm $\times$ 320nm. 

\subsection{Optical characterization} \label{sec:OpticalCharacterization}

\begin{figure}
    \centering
    \includegraphics[width=0.8\textwidth]{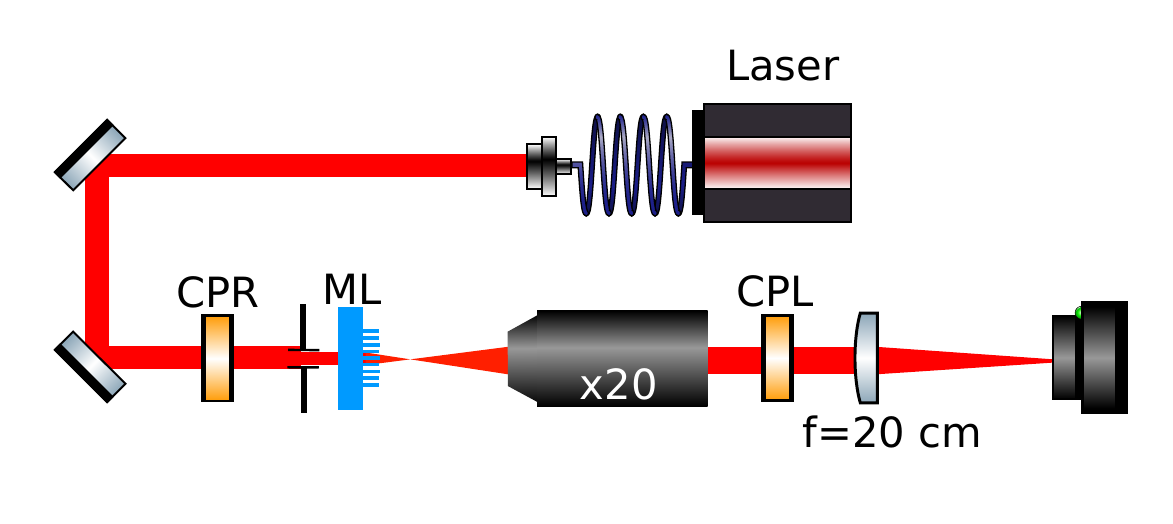}
    \caption{
    Optical setup used to characterize the metalenses. A collimated laser beam passes through a right handed circular polarizer (CPR), before going through an aperture with diameter 0.9~mm and then the metalens (ML). The beam is converted to left handed circularly polarized light and focused by the metalens. The resulting focal spot is imaged onto an IR camera using a x20 infinity corrected microscope objective and a planoconvex lens. A left handed circular polarizer (CPL) is placed in reverse between the microscope objective and the planoconvex lens, such that only the light which is converted from right to left handed circular polarization by the metalens is let through. When measuring the focal spot for the aspherical lens, the right handed circular polarizer is moved in front of one of the alignment mirrors, such that the handedness is changed by the mirror and the light can pass through the left handed circular polarizer. The aperture is used to ensure the lenses have the same effective numerical aperture, and a powermeter is used to ensure the same amount of light is transmitted through the aperture for all measurements.}
    \label{fig:optical_setup}
\end{figure}

\begin{figure}
    \centering
     \begin{subfigure}[b]{0.45\textwidth}
         \centering
         \includegraphics[width=\textwidth]{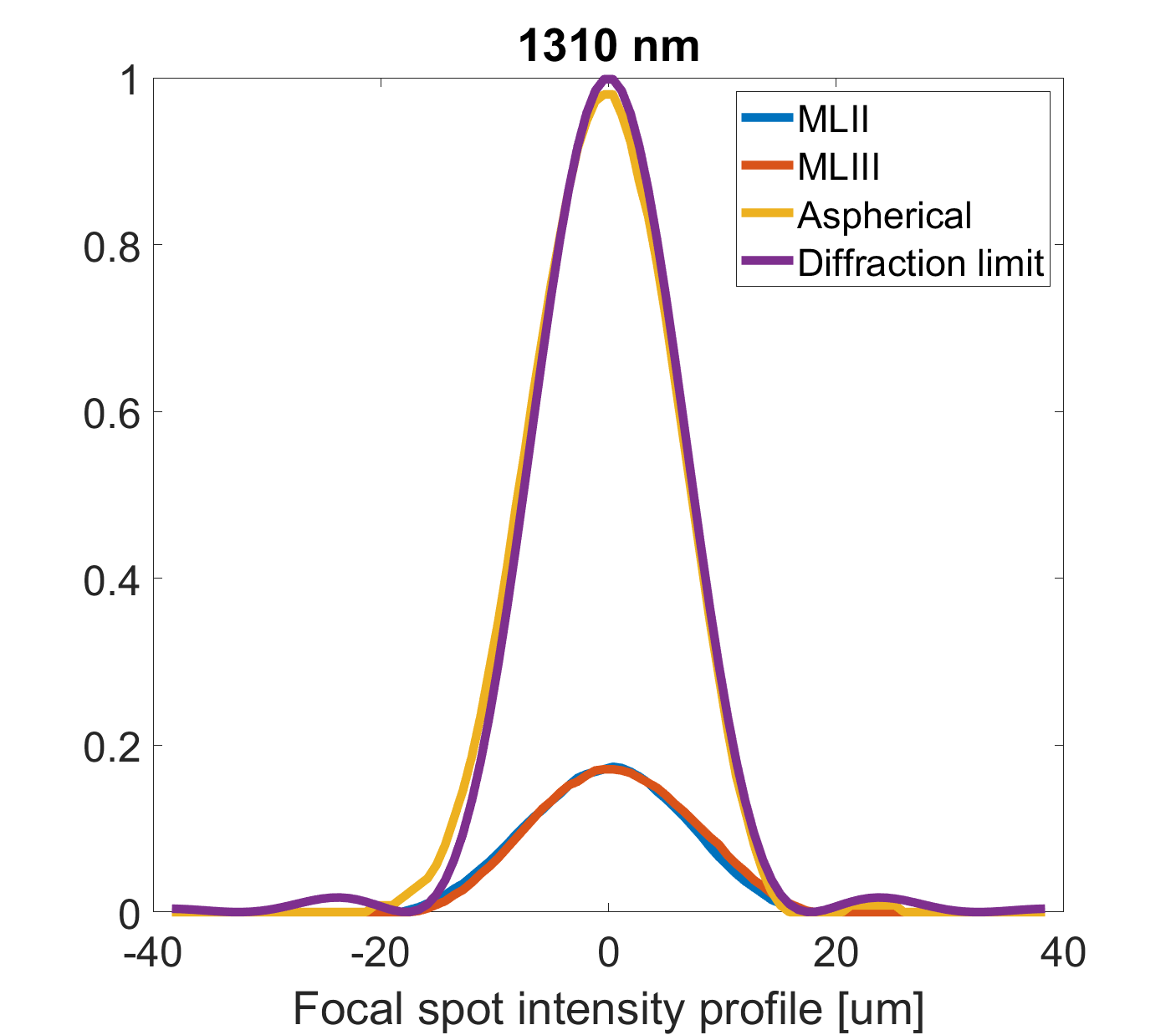}
         \caption{}
         \label{fig:focalspot_1310}
     \end{subfigure}
     \hfill
     \begin{subfigure}[b]{0.45\textwidth}
         \centering
         \includegraphics[width=\textwidth]{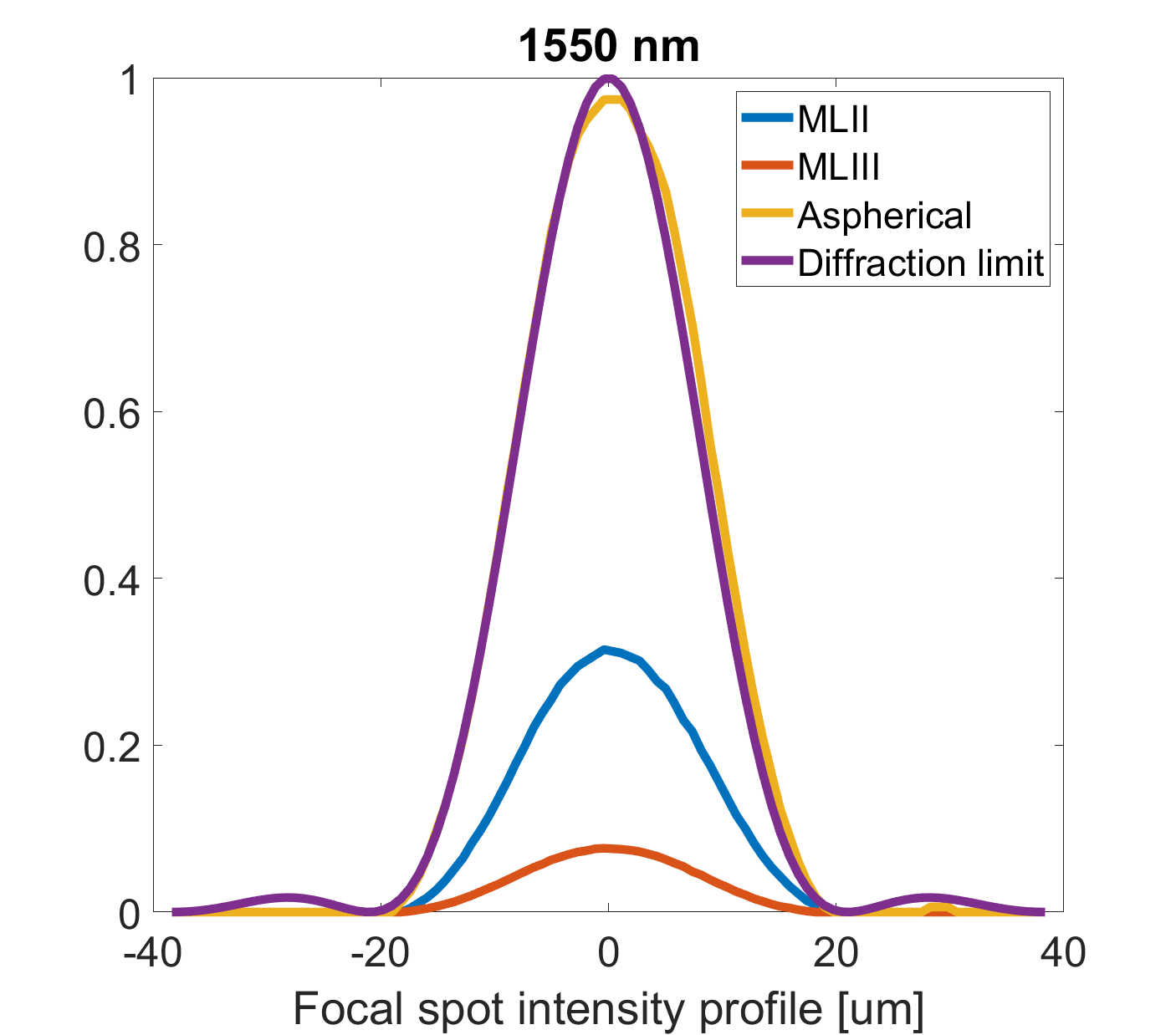}
         \caption{}
         \label{fig:focalspot_1550}
     \end{subfigure}
        \caption{Focal spot profiles measured for two metalenses and an anti-reflection coated aspherical lens when focusing a fully polarized and collimated laser beam of wavelength 1330 nm (a) and 1550 nm (b). For all lenses and for both wavelengths an aperture with diameter 0.9 mm has been placed in front of the lens.
        %In blue is a metalens with reduced effective dimensions due to deeper scallops, while in orange is a metalens with reduced dimensions resulting from oxidation and oxide stripping.
        %The curves are normalized to the peak intensity of the aspherical lens. Note that the metalens efficiencies have been divided by 0.7 to make up for the 30\% reflection from the back side of the Si substrate.
        }
        \label{fig:focalspot}
\end{figure}

\begin{figure}
    \centering
     \begin{subfigure}[b]{0.3\textwidth}
         \centering
         \includegraphics[width=\textwidth]{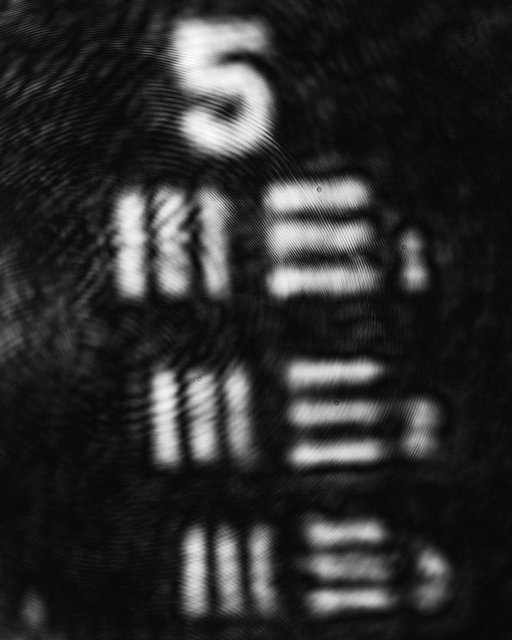}
         \caption{}
         \label{fig:aspherical_resolution_target_1550}
     \end{subfigure}
     \hfill
     \begin{subfigure}[b]{0.3\textwidth}
         \centering
         \includegraphics[width=\textwidth]{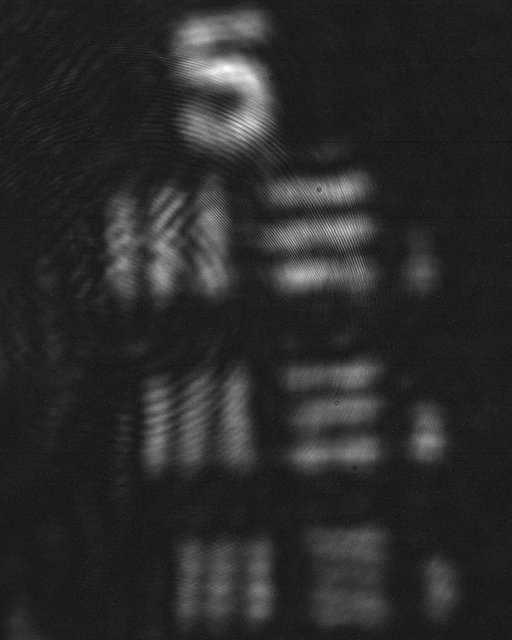}
         \caption{}
         \label{fig:RIE14_resolution_target_1550}
     \end{subfigure}
     \hfill
     \begin{subfigure}[b]{0.3\textwidth}
         \centering
         \includegraphics[width=\textwidth]{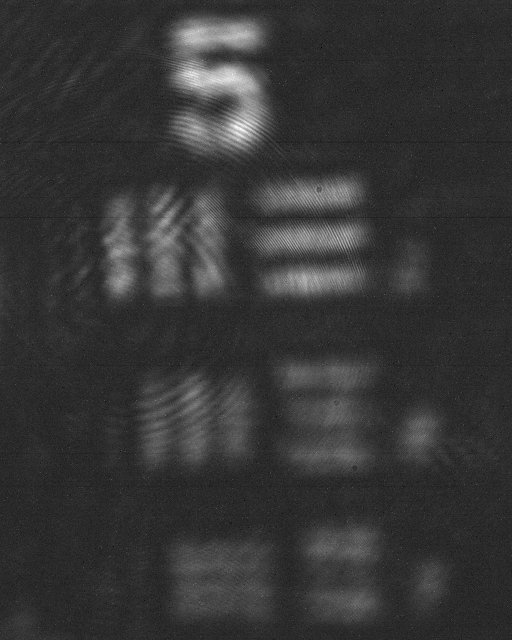}
         \caption{}
         \label{fig:RIE15_resolution_target_1550}
     \end{subfigure}
        \caption{Images of a resolution target using an aspherical lens (a), metalens MLII (b), and metalens MLIII (c). The groups of lines have widths 15.6 $\mu$m 13.9 $\mu$m and 12.4 $\mu$m. The target is illuminated by a 1550 nm laser beam, and to enable comparison of the lenses an aperture with diameter 0.9 mm has been placed in front of the lens. The contrast is significantly better for the aspherical lens, while the resolution is only slightly better - the 13.9 $\mu$m thick lines in (b) are resolved similarly to the 12.4 $\mu$m thick lines in (a).}
        \label{fig:resolution_target}
\end{figure}

The two metalenses from the fabrication steps discussed in Sec. \ref{sec:BoschEtch} were tested optically using the measurement setup shown in Fig.~\ref{fig:optical_setup}. One of the metalenses had comparatively large scallop depths of $\sim86$ nm and lateral dimensions on the order of 307nm $\times $460nm as shown in Fig.~\ref{fig:ComparisonDRIE}~(ii) (herafter called MLII). The other metalens had shallower scallops of $\sim29$ nm and lateral dimensions on the order of 210nm $\times$ 320nm, shown in Fig.~\ref{fig:ComparisonDRIE}~(iii) (hereafter referred to as MLIII). Note that while the metasurfaces were designed for left circular polarization, the transfer of the pattern to the nanoimprint master led to opposite rotation of cylinders compared to the simulations and consequently a change in the handedness of the circular polarized light for operation. Hence for the optical characterization, the metasurfaces are illuminated with right handed circular polarized light, and the metasurfaces focus the cross-polarized left handed circular polarized light. 

The focal spots of the metasurfaces are shown in Fig.~\ref{fig:focalspot} for two wavelengths,  $1.31\mu$m and $1.55\mu$m, together with the focal spot of an anti-reflection coated aspherical lens with the same focal length (10~mm) and the same aperture (0.9~mm diameter) for comparison. We see that both metalenses achieve diffraction limited focusing, having the same spot size as the aspherical lens matching the theoretical diffraction limit. As discussed in Sec. \ref{sec:PhysPrinc}, discrepancies from the target dimensions in the fabricated structures do not primarily affect the focal spot, but rather affect the lens efficiencies. MLII and MLIII ended up with effective dimensions that are smaller than the target dimensions of the optical design. At $1.31\mu$m both metalenses have a measured efficiency of 17\% compared to the aspherical lens, while at $1.55\mu$m MLII has a measured efficiency of 30\% and MLIII has a measured efficiency of 8\%. The measurements were made by comparing the peak values in Fig.~\ref{fig:focalspot}, and by disregarding reflection at the substrate back-side (by dividing the measured intensity by $0.7$) for comparison with the simulations in Sec. \ref{sec:Simulations}. We expect the efficiency values to increase as further process optimization (as discussed in Sec. \ref{sec:Discussion}) leads to better precision in reaching the target dimensions of the optical design. For incident light that has left handed circular polarization, the metalenses are divergent, having a focal length of -10~mm. This was confirmed by switching the polarizers and observing the virtual focal spot visible when bringing the metalens 10~mm inside the working distance of the microscope objective.

The same measurement setup was also used to take images of a resolution target using the metalenses.
For these measurements the resolution target was placed 2~cm in front of the metalens, and the image plane 2~cm behind the metalens was imaged onto the camera using the microscope objective and planoconvex lens.
Fig.~\ref{fig:resolution_target} shows the resulting images for the two metalenses and the aspherical lens using the 1550~nm laser. The aspherical lens has clearly better contrast owing to higher efficiency, while the resolution is only slightly better for the aspherical lens, since 13.9~$\mu$m thick lines are resolved similarly by the metalens as 12.4~$\mu$m thick lines are resolved by the aspherical lens.
Since all three lenses are observed to have the same diffraction limited focal spot size when the incoming light is collimated parallel to the optical axis, the slight difference in resolution when imaging the resolution target is likely due to coma~\cite{arbabi2016miniature}.

\section{Discussion} \label{sec:Discussion}
Our results have demonstrated the feasibility of using UV-NIL with subsequent continuous RIE and Bosch DRIE to fabricate diffraction limited metalenses. Further optimization towards high throughput production relevant processing should aim at improving resist adhesion upon stamp detachment and a reduction of resist broadening at the base of the resist pillars in order to obtain greater precision in reaching the target dimensions (and thereby raising the efficiency of the lenses). This section discusses these challenges in turn.

A significant challenge in our UV-NIL patterning process was to avoid delamination of the resist upon detachment of the soft-stamp. To some extent, our experiences seems to indicate a degree of trade-off between achieving a low residual layer thickess (RLT) in the resist (Sec. \ref{sec:NIL}) and its adhesion to the substrate. Despite the use of plasma-activation of cleaned substrates and quick subsequent application of an adhesion promoter (Sec. \ref{sec:NIL}), the soft-stamp was in need of replacement typically after 3-5 imprints due to delamination of resist into the holes of the stamp (see Fig. \ref{fig:Sketch_Softstamp} for sketch of the softstamp). This was in the case of the patterns of variant B in Table \ref{tab:LateralDimensions} which have the lowest filling factors $F=0.12$. For the case of variants A and C (which have larger filling factors of $F=0.17$ and $F=0.24$), the resist patterns more or less consistently delaminated on the first imprint. However, the issues with delamination seemed only to occur after having switched to the less viscous resist formulation (mr-NIL210-200nm) for which the desired low residual layer thickness (RLT) resist values were attained. While using the more viscous resist formulation (mr-NIL210-500nm) the imprint patterns of all filling factors more or less consistently turned out well. Unfortunately, as discussed in Sec. \ref{sec:NIL}, the resulting RLT values of the more viscous resist formulation were too large for the subsequent etching steps. Further process development of the UV-NIL patterning steps should therefore consider varying the RLT further by dilution of mr-NIL210-500nm and see if there exists a lower threshold of the RLT at which the adhesion issue ceases. The silicon pillars in our master wafer had slightly angled sidewalls ($>80^\text{o}$) which are known to facilitate the release of the imprint stamp \cite{schift2010nanoimprint}. Another strategy could be to test even larger angles: Tuning the etch properties of the master fabrication may allow for controllable sidewall angles, and for a systematic analysis of these with respect to soft stamp release. Although slight sidewall angles may be beneficial in this respect, the transfer of such sidewalls to the resist pillars in the mask may add uncertainty for reaching the desired lateral target dimensions.

The occurence of broadening at the base of the resist pillars (like "top-hats", see Fig. \ref{fig:ImprintmrNIL210}) lead to a broadening of the etched Si pillars in comparison to the mask dimensions of variant B in Table \ref{tab:LateralDimensions} (becoming roughly 180 nm too large). While we demonstrated that this could be compensated for by both increasing the lateral etch depth (i.e. scallop depths) of the Bosch pulsed DRIE (shown in Fig. \ref{fig:ComparisonDRIE}(ii)) and through oxidizing and stripping (shown in Fig. \ref{fig:ComparisonDRIE}(iii)) as discussed in Sec. \ref{sec:BoschEtch}, the addition of these processing steps also add uncertainties in predicting the resulting dimensions of the Si pillars: As was noted in Secs. \ref{sec:BoschEtch} and \ref{sec:OpticalCharacterization}, the resulting effective dimensions of the Si pillars with large scallop depths became slightly too small in comparison to the target values, which in turn may explain why the lens efficiencies are lower than their theoretical limits. Similar resist broadening to what we have observed seems to be commonly encountered in literature \cite{hamdana2018nanoindentation, si2017consecutive, plachetka2013tailored}. The authors of \cite{hamdana2018nanoindentation} demonstrate that the unwanted broadening at the base of the resist pillars can be successfully removed along with the RLT layer by use of an O$_2$ plasma, leaving the resist pillars with almost vertical walls. However, achieving similar results for rotating rectangular pillars where the minimum distance between pillars vary over the lens will likely require significant process development.

Both strategies of processing away the issues of resist broadening in the imprinted resist discussed so far may require significant process development in order to reduce tertainty in the resulting Si pillar dimensions. It would be preferable, therefore, to avoid the broadening in the first place: Avoiding the need to remove or correct for the resist "feet" at the base of the pillars, is expected to lead to better precision in reaching the target dimensions. This in turn should make it possible to develop more robust processes towards achieving diffraction limited metalenses of high efficiency. We believe the resist broadening likely originates from equivalent broadening being already present in master Si pillars and/or in the NIL stamp holes since the UV-cured resist generally follows the pattern of the stamp. It may be worth considering whether  process development of the NIL master fabrication can lead to Si pillar patterns without curvature at the base.

\section{Conclusion}
Diffraction limited dielectric metalenses have been fabricated using UV-Nano Imprint Lithography (UV-NIL) and a combination of continuous Reactive Ion Etching (RIE) and pulsed Bosch Deep Reactive Ion Etching (DRIE). These are standard silicon (Si) processing techniques that are relevant as the metasurface research field turns towards applications. In particular UV-NIL has been proposed as a strong candidate to replace the use of Electron Beam Lithography when seeking to achieve a high throughput and large area patterning technique. %Furthermore, UV-NIL and Bosch DRIE are techniques that can allow metasurfaces fabrication to be integrated into the same Si process lines that already are used for making e.g. detectors, thereby having the possibility of leading to significant cost savings and miniaturization. 
Simulations show that the "washboard-type" sidewall surface roughness characteristic of the Bosch DRIE process can be compensated for by increasing the lateral dimensions of Si pillars, and the fabricated structures have demonstrated diffraction-limited imaging despite the fact that its metastructure contains relatively large scallop depths. As such, the characteristic sidewall surface roughness of Bosch DRIE does not prevent the technique from being a strong candidate towards industrial metalens fabrication. It may however face some fundamental challenges in compensating for its sidewall roughness if seeking to fabricate nano-structures separated by high aspect ratio gaps.

The main challenges towards fabrication of the metalenses have been issues with delamination of the resist mask upon stamp removal and resist broadening at the base of the resist pillars. The latter lead to the lateral dimensions of the resulting Si pillars after etching being too large. These were compensated for by increasing the lateral etch depths in the pulsed Bosch Deep Reactive Ion Etching: I.e. the effective dimensions were reduced by increasing the scallop sizes. This resulted in well functioning diffraction limited lenses with measured efficiencies of 30\% and 17\% at wavelengths $\lambda=1.55\mu$m and $\lambda=1.31\mu$m, respectively. Process optimization strategies are discussed to improve resist adhesion and resolve the issue of resist broadening. The latter strategies should lead to improved precision in reaching the desired Si pillar dimensions, which in turn is expected to raise the efficiency of the lenses.

\section*{Funding}
The research leading to these results has received funding from EEA Grants 2014-2021, under Project contract no.5/2019.
% Please identify all appropriate funding sources by name and contract number. Funding information should be listed in a separate block preceding any acknowledgments. List only the funding agencies and any associated grants or project numbers, as shown in the example below:\\
% \\
% National Science Foundation (NSF) (1253236, 0868895, 1222301); Program 973 (2014AA014402); Natural National Science Foundation (NSFC) (123456).\\
% \\
% OSA participates in \href{https://www.crossref.org/fundingdata/}{Crossref's Funding Data}, a service that provides a standard way to report funding sources for published scholarly research. To ensure consistency, please enter any funding agencies and contract numbers from the Funding section in Prism during submission or revisions.

%\section*{Acknowledgments}

\section*{Disclosures}

% Disclosures should be listed in a separate nonnumbered section at the end of the manuscript. List the Disclosures codes identified on OSA's \href{http://www.osapublishing.org/submit/review/conflicts-interest-policy.cfm}{Conflict of Interest policy page}, as shown in the examples below:

% \medskip

% \noindent ABC: 123 Corporation (I,E,P), DEF: 456 Corporation (R,S). GHI: 789 Corporation (C).

% \medskip

% \noindent If there are no disclosures, then list ``
The authors declare no conflicts of interest.

\bibliography{ChrisBibliography.bib}

%%%%%%%%%% If preparing manually:
% \begin{thebibliography}{1}
% \newcommand{\enquote}[1]{``#1''}

% \bibitem{Zhang:14}
% Y.~Zhang, S.~Qiao, L.~Sun, Q.~W. Shi, W.~Huang, L.~Li, and Z.~Yang,
%   \enquote{Photoinduced active terahertz metamaterials with nanostructured
%   vanadium dioxide film deposited by sol-gel method,}
%   {\protect\JournalTitle{Optics Express}} \textbf{22}, 11070--11078 (2014).

% \bibitem{OSA}
% {Optical Society}, \enquote{{OSA Publishing},}
%   \url{http://www.osapublishing.org}.

% \bibitem{FORSTER2007}
% P.~Forster, V.~Ramaswamy, P.~Artaxo, T.~Bernsten, R.~Betts, D.~Fahey,
%   J.~Haywood, J.~Lean, D.~Lowe, G.~Myhre, J.~Nganga, R.~Prinn, G.~Raga,
%   M.~Schulz, and R.~V. Dorland, \enquote{Changes in atmospheric consituents and
%   in radiative forcing,} in \enquote{Climate Change 2007: The Physical Science
%   Basis. Contribution of Working Group 1 to the Fourth assesment report of
%   Intergovernmental Panel on Climate Change,}  S.~Solomon, D.~Qin, M.~Manning,
%   Z.~Chen, M.~Marquis, K.~B. Averyt, M.~Tignor, and H.~L. Miler, eds.
%   (Cambridge University Press, 2007).

% \end{thebibliography}

\end{document}